\documentclass{article}

\usepackage[numbers]{natbib}         % bibliography style
\usepackage[colorlinks]{hyperref}    % better urls in bibliography
\usepackage[english]{babel} %%% 'french', 'german', 'spanish', 'danish', etc.
\usepackage{amssymb}
\usepackage{amsmath}
\usepackage{txfonts}
\usepackage{mathdots}
\usepackage[classicReIm]{kpfonts}
\usepackage[dvips]{graphicx} %%% use 'pdftex' instead of 'dvips' for PDF output
\usepackage[a4paper, portrait, margin=1in]{geometry}
\usepackage{tabularx}
\usepackage{longtable}
\begin{document}

%\selectlanguage{english} %%% remove comment delimiter ('%') and select language if required

\textbf{\Large Machine Learning in Quantitative PET Imaging}

\textbf{ }

Tonghe Wang${}^{1,2}$, Yang Lei${}^{1}$, Yabo Fu${}^{1}$, Walter J. Curran${}^{1,2}$, Tian Liu${}^{1,2}$ and Xiaofeng Yang${}^{1,2}$*

${}^{1}$Department of Radiation Oncology, Emory University, Atlanta, GA 

${}^{2}$Winship Cancer Institute, Emory University, Atlanta, GA

\noindent 
\bigbreak
\bigbreak
\bigbreak

\textbf{Corresponding author: }

Xiaofeng Yang, PhD

Department of Radiation Oncology

Emory University School of Medicine

1365 Clifton Road NE

Atlanta, GA 30322

E-mail: xiaofeng.yang@emory.edu

\bigbreak
\bigbreak
\bigbreak
\bigbreak
\bigbreak
\bigbreak

\textbf{Abstract}

This paper reviewed the machine learning-based studies for quantitative positron emission tomography (PET). Specifically, we summarized the recent developments of machine learning-based methods in PET attenuation correction and low-count PET reconstruction by listing and comparing the proposed methods, study designs and reported performances of the current published studies with brief discussion on representative studies. The contributions and challenges among the reviewed studies were summarized and highlighted in the discussion part followed by.

\noindent \eject 

\noindent 
\section{ INTRODUCTION}

Positron emission tomography (PET) has been used as a non-invasive functional imaging modality with a wide range of clinical applications. By providing the information of metabolic processes in human body, it is utilized for various purposes including staging tumor and finding metastases in oncology,\cite{RN1743, RN1742, RN1746, RN1740, RN1745, RN1738, RN1741, RN1739, RN1744} gross target volume definition in radiation oncology,\cite{RN1755, RN1757, RN1756} myocardial perfusion imaging in cardiology,\cite{RN1754, RN1747} and complex neurological disorders investigation.\cite{RN1758} Among these applications, the accuracy of tracer uptake quantification has been less recognized than other characteristics of PET such as sensitivity. Recently, with the focus shifting towards precision medicine, it is of great clinical interest in accurate quantitative measurement on tracer uptake, where the role of PET is expanded to more demanding applications such as estimation of target occupancy in drug development,\cite{RN1748} therapeutic response monitoring,\cite{RN1749, RN1750, RN1752} and treatment outcome prediction as a prognostic factor.\cite{RN1751}
The accuracy of the uptake quantification, on which these potential applications rely, would be limited due in part to the physical nonidealities including photon attenuation\cite{RN1759}, low count statistics,\cite{RN1760, RN1761} partial-volume effect and etc.\cite{RN1763, RN1753, RN1762}  These issues result in bias, uncertainty and artifacts on the PET images, which would degrade the utility of both qualitative and quantitative assessments. They are becoming conspicuous with the introduction of advanced PET scanners and applications in the recent years, such as MR-combined PET (PET/MR), low-count scanning protocol and high-resolution PET. These novel PET technologies aim to incorporate anatomical imaging modality for better soft tissue visualization, reduce administered activity and shorten scan time of conventional PET and CT-combined PET (PET/CT), and increase detection capability on radiopharmaceutical accumulation in structures of millimeter size, all of which are highly desirable for clinical practice. On the other hand, unconventional or suboptimal scan schemes are proposed to achieve these clinical benefits, which brings new technical challenges that have not been seen in conventional PET in implementing these features without compromising the quantification capabilities.

Many methods have been proposed to deal with these inherent deficiencies of PET since it was invented, and are still being improved to tackle the difficulties emerged from these novel implementations. For example, before PET/CT was widely implemented, transmission scan with an external positron source rotated around the patient was used to determine the attenuation of patient body for attenuation correction (AC). The additionally acquired CT images in PET/CT is used to derive the 511keV linear attenuation coefficient maps to model photon attenuation by a piecewise linear scaling algorithm, and to provide high-resolution anatomical images.\cite{RN1167, RN1759} In recent years, the incorporation of magnetic resonance (MR) imaging with PET (PET/MR) becomes a promising alternative to the existing PET/CT system by providing excellent soft tissue visualization without ionizing radiation. However, the MR voxel intensity is related to proton density rather than electron density, thus it cannot be directly converted to 511 keV attenuation coefficient for AC process. Conventional methods propose to assign piecewise constant attenuation coefficients on MR images based on the segmentation of tissues.\cite{RN1865, RN1864} The segmentation can be done by either manually-drawn contours\cite{RN1169} or automatic classification methods.\cite{RN1170, RN1171, RN1172} However, these methods are limited by misclassification and inaccurate prediction of bone and air regions caused by their ambiguous relations in MR voxel intensities. Instead of segmentation, other methods were used to warp atlases of MR images labeled with known attenuation to patient-specific MR images by deformable registration or pattern recognition, but their efficacy highly depends on the performance of the registration. Moreover, the atlases that are usually created on normal anatomy could not represent anatomic abnormalities in clinical practice.\cite{RN1174, RN1173}

Low-count PET protocol aims to reduce administrated activity, which is attractive for it clinical utility.\cite{RN1845} The reduced ionization radiation is desirable to pediatric PET scans since the accumulated imaging dose can be a big concern for pediatric patients who are more sensitive to radiation and have longer life expectancy than adults.\cite{RN1812, RN1706} It would also help reduce radiation exposure to patients who undergo radiation therapy and have multiple serial PET scans from pretreatment until the completion of radiation treatment for radiotherapy response evaluation.\cite{RN1835, RN1834} The shortened scan time is beneficial for motion control as well as can potentially increase patient load. More importantly, in dynamic PET imaging that scans a particular bed position over a sequence of time frames, the counts at each frame is much lower than that in static PET imaging. Dynamic PET imaging aims to provide a voxel-wise parametric analysis for kinetic modeling, the accuracy of which highly depends on the image quality.\cite{RN1859} However, the low count statistics would result in increased image noise, reduced contrast-to-noise ratio, and large bias in uptake measurement.\cite{RN1765} Both hardware-based and software-based solutions have been proposed for the low-count PET scanning. Advances in PET instruments such as lutetium-based detectors, silicon photomultipliers and time-of-flight compatible scanners are able to considerably increase the acquisition efficiency thus lower the measurement uncertainty. \cite{RN1767, RN1766} Meanwhile, software-based post-processing and the usage of noise regularization in reconstruction would penalize the differences among neighboring pixels  in order to have a smooth appearence.\cite{RN1769, RN1770, RN1768}

As reviewed above, hardware advances, mainly focusing on the development in the scintillator, photodetector and electronics for high spatial/energy/timing resolution performance, would directly raise the fundamental capability of the entire PET system. However, it heavily relies on the breakthrough of basic research in material science and nuclear electronics, and usually requires intensive labor for its setup and commissioning. After decades of continuous optimization of PET systems, further hardware improvements are difficult and expected to be costly. On the other hand, software improvements, particularly novel image processing and reconstruction algorithms, are complementary to hardware improvements and easier to implement at a lower cost.

Inspired by the rapid expansion of artificial intelligence in both industry and academia in recent years, many research groups have attempted to integrate machine learning-based methods into medical imaging and radiation therapy.\cite{RN1781, RN1782, RN1780, RN1783, RN1779} The common machine learning applications include detection, segmentation, characterization, reconstruction, registration and synthesis.\cite{RN1717, RN1855, RN1810, RN1794, RN1715, RN1795, RN1708} Before the permeation of artificial intelligence into these sub-fields, conventional imaging processing methods have been developed for decades. The conventional algorithms are usually very different from each other in many aspects such as workflow, assumption, complexity, requirement on prior knowledge and etc, and the detailed implementation highly depends on the task.\cite{RN1364, RN726, RN722, RN1143, RN1119} Compared with conventional methods, machine learning-based methods share a general framework using a data-driven approach. Supervised learning workflow usually consists of a training stage, i.e. a machine learning model is trained by the training datasets to find the patterns between the input and its training target, and a predication stage, where the trained model maps an input to an output. Unlike the conventional methods whose performance is usually sensitive to its hyperparameter settings, machine-learning based methods are more robust, whose effectiveness largely depends on the representativeness of the training datasets used.
In this paper, we reviewed the emerging machine learning-based methods for quantitative PET imaging. The PET scans in these reviewed literatures used glucose analogue 2-18F-fluoro-2-deoxy-D-glucose (FDG) for uptake if not explicitly stated otherwise. Specifically, we reviewed the general workflows and summarized the recently published studies of learning-based methods in dealing with PET AC and low-count PET reconstruction in the following sections respectively, and briefly introduced the representative ones among them with a discussion on the trend and future direction in the end of each section.

\noindent 
\section{PET AC}

As mentioned in the introduction, CT is able to provide the attenuation coefficient maps to correct the loss of annihilation photons due to attenuation process in the patient body. However, CT is unavailable in a PET-only scanner or a PET/MR scanner which does not provide photon attenuation coefficients. Current studies proposed several schemes to address this issue, and machine learning is involved in different forms with different purposes. The general workflows can be divided into two groups depending on whether anatomical image is acquired (MR in PET/MR) or not (PET-only scanner). Most proposed methods were implemented on PET brain scans, with a few on pelvis and whole body. The ground truth for training and evaluation was PET with AC by CT images. Relative bias in percentage from ground truth in selected volumes-of-interest (VOIs) was usually reported to evaluate the quantification accuracy. Several studies compared their proposed methods with conventional MR-based PET AC methods such as atlas-based methods and segmentation-based methods.

\noindent 
\subsection{Synthetic CT for PET/MR}

For a PET/MRI scanner, the most common strategy is to generate synthetic CT (sCT) images from MR images, and replace CT by sCT in the PET AC procedure. The synthesis among images of different modalities has been developed in recent years.\cite{RN1790, RN1793, RN1711, RN1705, RN1685, RN1809, RN1684, RN1811, RN1714, RN1806, RN1807, RN1800, RN1789, RN1710, RN1712, RN1689, RN1709, RN1679} Many studies investigated the feasibility of using sCT for PET AC in brain and whole body imaging, which are summarized in Table I. The sCT was generated by assigning CT numbers to corresponding tissue labels segmented on MR as proposed by Liu \textit{et al.}\cite{RN1813} They used convolutional auto-encoder (CAE), which consists of a connected encoder and decoder network, to generate CT tissue labels from T1-weighted MR images. The encoder and decoder probe the image features and reconstruct piecewise tissue labels, respectively. They reported an average bias of -0.7$\mathrm{\pm}$1.1 \% among all selected VOIs in brain in the PET AC using their sCT, which is significantly better in most of the VOIs than two comparing methods, Dixon-based soft-tissue/air segmentation and anatomic CT-based template registration. The limitation of this strategy is its requirement of labelling on training datasets and small number of classified tissue types.
sCT can also be generated through a direct mapping from MR. Gaussian mixture regression,\cite{RN1823} support vector regression\cite{RN1824} and random forest\cite{RN1808} have been used before deep learning was implemented. For example, Yang \textit{et al.} proposed a random forest-based method to train a set of decision trees. Each decision tree learns the optimal way to separate a set of training paired MRI and CT patches into smaller and smaller subsets to predict the CT intensity. When a new MRI patch is put into the model, the sCT intensity is estimated as the combination of the predicted results of all decision trees. They used a sequence of alternating random forests under the framework of an iterative refinement model  to consider both the global loss of training model and the uncertainty of training data falling into child nodes, with combination of discriminative feature selection. \cite{RN1808} The authors reported good visual agreement of brain AC PET with bias ranging from -1.61\% to 3.67\% among all selected regions. Since deep learning methods have been proved to be successful in image style transfer in computer vision field, sCT from MR can be generated by many off-the-shelf algorithms. One of the most popular networks is Unet and its variant. These convolutional neural network (CNN)-based methods aim to minimize the loss function that includes the pixel-intensity and pixel-gradient difference between training target CT and generated sCT. The implementation and results of Unet in PET AC has been reported in ref \cite{RN1815} for brain, ref \cite{RN1821, RN1827} for pelvis, ref \cite{RN1817} for pediatric brain tumor patients and ref \cite{RN1825} for neuroimaging with 11C-labeled tracers. In order to incorporate multiple MR images from different sequences as input, Gong \textit{et al.} modified Unet by replacing the traditional convolution module by group convolutional module to preserve the network capacity while restricting the network complexity.\cite{RN1816} This modification was demonstrated to successfully lower the systematic bias and errors in all regions of a conventional Unet. 
Apart from CNN-based methods, generative adversarial networks (GAN) has also been exploited in generating sCT for PET AC. GAN has a generative network and a discriminative network that are trained simultaneously. With MR as input and sCT as output, the discriminative network distinguishes the sCT from training CT images. It can be formulated as solving a minimization problem of the discriminative loss, in addition to pixel-intensity loss and pixel-gradient loss to improve sCT image quality. Arabi \textit{et al.} proposed a deep learning adversarial semantic structure that combined a synthesis GAN and a segmentation GAN.\cite{RN1818} The segmentation GAN segmented the generated sCT into air cavities, soft tissue, bone and background air such that it can regularize the sCT generation process by back-propagating gradients. The proposed method was reported a mean bias less than 4\% in selected VOIs in brain PET, which was slightly worse than atlas-based methods (up to 3\%) but better than segmentation-based methods (up to -10\%).

\noindent 
\subsection{PET AC for PET-only}

Although PET-only scanner does not provide anatomical images such as MR, it has been shown that the non attenuation-corrected (NAC) PET could be used to generate sCT by the powerful image style transfer ability of deep learning methods. Similar to the sCT generation from MR, the sCT images were generated from the NAC PET images using the machine learning model trained by pairs of NAC PET and CT images that were acquired from a PET/CT scanner. Again, Unet and GAN were the two common networks adopted for this application.\cite{RN1819, RN1703, RN1822, RN1814} Among these studies, learning-based PET AC on whole body PET imaging was investigated by Dong \textit{et al.} for the first time.\cite{RN1826} A cycle-consistent GAN (CycleGAN) method that combined a self-attention Unet for generator architecture and a fully convolutional network for discriminator architecture was employed. The method learns a transformation that minimizes the difference between sCT, generated from NAC PET, and true CT. It also learns an inverse transformation such that cycle NAC PET image generated from the sCT is close to true NAC PET image. Self-attention strategy was utilized to identify the most informative component and mitigate the disturbance of noise. The average bias of the proposed method on selected organs and lesions ranged from -1.06\% to 3.57\% except for lungs (10.72\%).
The above strategy still requires the step of PET reconstruction using the sCT in addition to the mapping step. Thus, attempts were made to directly map the AC PET from the NAC PET images by exploiting the deep learning methods to bypass the PET reconstruction step. Yang \textit{et al.} and Shiri \textit{et al.} both proposed Unet-based methods in direct brain PET AC and demonstrated the feasibility.\cite{RN1829, RN1828} Dong \textit{et al.} again investigated the whole body PET direct AC using a supervised 3D patch-based CycleGAN.\cite{RN1826} The CycleGAN has a NAC-to-AC PET mapping and an inverse AC-to-NAC PET mapping in order to constrain the NAC-to-AC mapping to approach a one-to-one mapping. Since NAC PET images have similar anatomical structures to the AC PET images but lack contrast information, residual blocks were integrated into the network to learn the differences between NAC PET and AC PET. They reported the average bias of the proposed method on selected organs and lesions ranging from 2.11\% to 3.02\% except for lungs (-17.02\%). When comparing with the results of Unet and GAN training and testing on the same datasets, the proposed CycleGAN method achieves a superior performance in most evaluation metrics and much less bias in lesions.

\noindent 
\subsection{Discussion}

Although it is difficult to specify the tolerance level of quantification error before it affect clinician’s judgement, the general consensus is that quantification errors of 10\% or less typically do not affect diagnosis.\cite{RN1174} Based on the average relative bias represented by these studies, almost all of the proposed methods in the studies met this criterion. However, it should be noted that due to the variation among study objects, the bias in some VOIs of some patients may exceed 10\%.\cite{RN1819, RN1815, RN1826, RN1703, RN1822, RN1821, RN1828} It suggests that special attention should be given to the standard deviation of bias as well as its mean when interpreting results since the proposed methods may have poor local performance that would affect some patients. On the other hand, listing or plotting the results of every data points, or at least the range, instead of simply giving a mean$\mathrm{\pm}$STD in presenting results, would be more informative in demonstrating the performance of the proposed methods.

Various learning-based approaches have been proposed in the reviewed studies. However, the reported errors among these studies cannot be fairly compared because of different datasets, evaluation methodology, and reconstruction parameters. Thus, it is impracticable to conclude the best method in performance based on these reported results from different studies. Some studies compared their proposed methods with other competing methods using same datasets, which may reveal the advantage and limitation of the methods selected in comparison. For example, a lot of studies compared their proposed method with conventional segmentation-based or atlas-based methods, and almost all the learning-based methods gain the upper hand in less bias on average with less variation among patients and VOIs in study, which indicates the advantage of the data-driven approaches over model-based methods. The comparison among the learning-based methods, however, is much less common in current studies, probably because these methods are just published in recent two years. As mentioned above, Dong \textit{et al.} compared the CycleGAN, GAN and Unet in direct NAC PET-AC PET mapping on whole body PET images, and demonstrated the superiority of CycleGAN over other two approaches owing to the addition of inverse mapping from AC PET to NAC PET that degenerates the ill-posed interconversion between AC PET and NAC PET to be a one-to-one mapping.\cite{RN1826}

Among the above PET/MR studies, different types of MR sequences have been adopted for sCT generation. The specific MR sequence used in each study usually depends on the accessibility. The optimal sequence yielding the best PET AC performance has not been studied. T1-weighted and T2-weighted sequences are the two of the most common MR sequences used in diagnosis. Due to their wide availability, sCT model can be trained from a relatively large number of datasets with CT and co-registered T1- or T2-weighted MR images regardless of PET acquisition. Thus, they have been utilized in multiple studies where the MR images in training datasets are not acquired with PET. \cite{RN1818, RN1823, RN1813, RN1825, RN1808} However, air and bone have little contrast in either T1- or T2-weighted MR images, which may impede the extraction of the corresponding features in learning-based methods. Two-point Dixon sequence can separate water and fat, which is suitable for segmentation. It has already been applied in commercial PET/MR scanner with combination of volume-interpolated breath-hold examination (VIBE) for Dixon-based soft-tissue and air segmentation for PET AC as a clinical standard.\cite{RN1830, RN1831} Its drawback is again the poor contrast of bone, which results in the misclassification of bone as fat. Learning-based approaches have been employed on Dixon MR images by Torrado-Carvajal et al.\cite{RN1827} They compared the performance of a Unet-based method with the vendor-provided segmentation based method on the same datasets. The relative bias of the proposed Unet-based method is 0.27$\mathrm{\pm}$2.59\%, -0.03$\mathrm{\pm}$2.98\% and -0.95$\mathrm{\pm}$5.09\% for fat, soft-tissue and bone, respectively, while it is 1.48$\mathrm{\pm}$6.51\%, -0.34$\mathrm{\pm}$10.00\% and -25.1$\mathrm{\pm}$12.71\% for segmentation-based methods. The learning-based methods significantly reduced the bias in bone with a much less standard deviation of all three VOIs among all patients, which indicates its superior capability in PET AC due to improved bone identification. In order to enhance the bone contrast to facilitate the feature extraction in learning-based methods, ultrashort echo time (UTE)– and/or zero echo time (ZTE) MR sequences have been recently highlighted due to its capability to generate positive image contrast from bone.\cite{RN1813} Ladefoged \textit{et al.} and Blanc-Durand \textit{et al.} demonstrated the feasibility of UTE and ZTE MR sequences using Unet in PET/MR AC, repectively.\cite{RN1815, RN1817} The former study reported -0.1\% (95\%CI: -0.2 to 0.5\%) bias on tumor without statistical significance from ground truth using Unet on UTE, which was superior to a significant bias of 2.2\% (95\%CI: 1.5 to 2.8\%) using the vendor-provided segmentation-based method on Dixon. The latter study reported the bias of Unet AC method on ZTE with a mean of -0.2\% and ranging from 1.7\% in vertex to -1.8\% in temporal lobe, and compared with a vendor-provided segmentation-based AC method on ZTE\cite{RN1832} which underestimated in most VOIs with a mean bias of -2.2\% and ranged from 0.1 to -4.5. It should be noted that neither of the studies compared the using of UTE/ZTE or conventional MR sequence under the same deep learning network. Thus, the advantage of this specialized sequence has not been validated. Moreover, compared with conventional T1-/T2-weighted MR images, the UTE/ZTE MR images have little diagnostic value on soft tissue while have a long acquisition time, which may hinder its utility in time-sensitive cases such as whole-body PET/MR scans. Other studies attempted to use multiple MR sequences as input in training and sCT generation since it is believed to be superior to single MR sequence in sCT accuracy. \cite{RN1816, RN1823, RN1821, RN1824} The most common combination is UTE/ZTE and Dixon, which provide contrast of bone against air and fat against soft tissue, respectively. However, so far there is no study that has quantitatively investigated the additional gain in PET AC accuracy from the multi-sequence input when comparing with single-sequence input. Thus, the necessity of the addition of MR sequence for PET AC needs to be further evaluated and to be balanced with the extra acquisition time.

In the above PET/MR studies, the CT and MR images in the training datasets were acquired separately on different machines. The reviewed learning-based sCT methods all require image registration between the CT and MR to create CT-MR pairs for training. For the machine learning-based sCT methods such as random forest, the performance is sensitive to the registration error in the training pairs due to its one-to-one mapping strategy. Deep learning methods such as Unet and GAN-based methods are also susceptible to registration error if using a pixel-to-pixel loss. Kazemifar \textit{et al.} showed that to the use of mutual information as the loss function in the generator of GAN can bypass the registration step in the training.\cite{RN1833} CycleGAN-based methods feature higher robustness to registration error since it introduces cycle consistence loss to enforce the structural consistency between original one and cycle one, (e.g., force cycle MRI generated from synthetic CT to be the same as original  MRI).\cite{RN1790, RN1677, RN1707, RN1711} These techniques helping reducing the requirement on registration accuracy have been proposed in the MR-sCT generation studies, and are worth exploration in the context of PET AC.

Whole-body PET scan has been an important imaging modality in finding tumor metastasis. Almost all of the reviewed studies developed their proposed methods for applications for brain imaging. Although in learning-based methods which are data-driven, the network and architecture are not designed for a specific site, the feasibility accepted for brain images may not be guaranteed for the cases of whole body due to the high anatomical heterogeneities and intersubject variability. The only two studies on learning-based whole-body PET AC are from Dong \textit{et al.} who proposed the CycleGAN-based method and investigated both the sCT generation strategy and direct mapping strategy.\cite{RN1826, RN1703} They reported average bias within 5\% in all selected organs except >10\% in lungs in both studies. The authors attributed the poor performance on lung to the tissue inhomogeneity and insufficient representative training datasets. Both studies are performed for PET-only scanner, and so far, there is no learning-based methods developed for PET/MR whole body scanner. Compared with PET-only scheme, the PET/MR provides the anatomical structural information from MR, while the integration of the additional MR into PET AC can be more challenging than brain scans since the MR may have a limited field of view (FOV), longer scan time that introduces more motion, and degraded image quality due to larger inhomogeneous-field region.

\noindent

\noindent Table I. Overview of learning-based PET AC methods.

\begin{longtable}{|p{1.2in}|p{0.8in}|p{0.4in}|p{1.0in}|p{0.7in}|p{0.7in}|} \hline 
Methods and strategy & PET or PET/MR & Site & \# of patients in training/testing datasets & Reported bias${}^{+}$ (\%) & Authors \\ \hline 
Deep convolutional auto-encoder (CAE) network\newline MR-$\mathrm{>}$tissue class (bone, soft-tissue, air) & PET/MR(3T Post-contrast T1-weighted) & Brain & 30 train/ 10 test & -0.7$\mathrm{\pm}$1.1 (-3.2, 0.4) among 23 VOIs & Liu \textit{et al.} 2018\cite{RN1813} \\ \hline 
Gaussian mixture regression\newline MR-$\mathrm{>}$sCT & PET/MR(T2-weighted+ultrashort echo time) & Brain & N.A.*/9 test & -1.9$\mathrm{\pm}$4.1 (-61, 34) in global & Larsson \textit{et al.} 2013\cite{RN1823} \\ \hline 
Support Vector Regression\newline MR-$\mathrm{>}$sCT & PET/MR(UTE and Dixon-VIBE) & Brain & 5 train/ 5 test & 2.16$\mathrm{\pm}$1.77 (1.32, 3.45) among 13 VOIs & Navalpakkam \textit{et al.} 2013\cite{RN1824} \\ \hline 
Alternating random forests\newline MR-$\mathrm{>}$sCT  & PET/MR(T1-weighted MP-RAGE) & Brain & 17 leave-one-out & (-1.61, 3.67) among 11 VOIs & Yang \textit{et al.} 2019\cite{RN1808} \\ \hline 
Unet\newline MR-$\mathrm{>}$sCT & PET/MR(ZTE)\newline  & Brain & 23 train/ 47 test & -0.2(-1.8, 1.7) among 70 VOIs & Blanc-Durand \textit{et al.} 2019\cite{RN1815} \\ \hline 
Unet\newline MR-$\mathrm{>}$sCT & PET/MR\newline 3T ZTE and Dixon & Pelvis & 10 train/ 26 test & -1.11$\mathrm{\pm}$2.62 in global & Leynes \textit{et al.} 2017\cite{RN1821} \\ \hline 
Unet\newline MR-$\mathrm{>}$sCT & PET/MR\newline Dixon & Pelvis & N.A./19 patients with 28 scans in test. & -0.95$\mathrm{\pm}$5.09 in bone\newline -0.03$\mathrm{\pm}$2.98 in soft tissue\newline 0.27$\mathrm{\pm}$2.59 in fat & Torrado-Carvajal et al 2019\cite{RN1827} \\ \hline 
Unet\newline MR-$\mathrm{>}$sCT & PET/MR(1.5T T1-weighted) & Brain & 44 train/ 11 validation/ 11 test & -0.49$\mathrm{\pm}$1.7 ${}^{11}$C-WAY-100635 PET\newline -1.52$\mathrm{\pm}$0.73 ${}^{11}$C-DASB\newline in global & Spuhler \textit{et al.} 2019\cite{RN1825} \\ \hline 
Unet\newline MR-$\mathrm{>}$sCT & PET/MR(ZTE and Dixon) & Brain & 14 leave-two-out & Absolute error (1.5\%-2.8\%) among 8 VOIs & Gong \textit{et al.} 2018\cite{RN1816} \\ \hline 
Unet\newline MR-$\mathrm{>}$sCT & PET/MR(UTE) & Brain & 79 (pediatric) 4-fold cross validation & -0.1(-0.2, 0.5) in 95\%CI among all tumor volumes & Ladefoged \textit{et al.} 2019\cite{RN1817} \\ \hline 
Deep learning adversarial semantic structure (DL-AdvSS)\newline MR-$\mathrm{>}$sCT & PET/MR(3T T1 MP-RAGE) & Brain & 40 2-fold cross validation & $\mathrm{<}$4 Among 63 VOIs & Arabi \textit{et al.} 2019\cite{RN1818} \\ \hline 
Hybrid of CAE and Unet\newline NAC PET-$\mathrm{>}$sCT-$\mathrm{>}$AC PET & PET(${}^{18}$F-FP-CIT) & Brain & 40 5-fold cross validation  & About (-8, -4) among 4 VOIs & Hwang \textit{et al.} 2018\cite{RN1822} \\ \hline 
Unet\newline NAC PET-$\mathrm{>}$ sCT-$\mathrm{>}$AC PET & PET & Brain & 100 train/ 28 test & -0.64$\mathrm{\pm}$1.99 (-4.18, 2.22) among 21 regions & Liu \textit{et al.} 2018\cite{RN1814}  \\ \hline 
Generative adversarial networks (GAN)\newline NAC PET-$\mathrm{>}$sCT-$\mathrm{>}$AC PET & PET & Brain & 50 train/ 40 test & (-2.5, 0.6) among 7 VOIs & Armanious \textit{et al.} 2019\cite{RN1819} \\ \hline 
Cycle-consistent GAN\newline NAC PET-$\mathrm{>}$sCT-$\mathrm{>}$AC PET & PET & Whole body & 80 train/ 39 test & (-1.06,10.72) among 7 VOIs\newline 1.07$\mathrm{\pm}$9.01 in lesion & Dong \textit{et al.} 2019\cite{RN1703} \\ \hline 
Unet\newline NAC PET-$\mathrm{>}$AC PET & PET & Brain & 25 train/ 10 test & 4.0$\mathrm{\pm}$15.4 among 116 VOIs & Yang \textit{et al.} 2019\cite{RN1828} \\ \hline 
Unet\newline NAC PET-$\mathrm{>}$AC PET & PET & Brain & 91 train/ 18 test & -0.10$\mathrm{\pm}$2.14 among 83 VOIs & Shiri \textit{et al.} 2019\cite{RN1829} \\ \hline 
Cycle-consistent Generative adversarial networks (GAN)\newline NAC PET-$\mathrm{>}$AC PET & PET & Whole body & 25 leave-one-out+\newline  10 patients*3 sequential scan tests & (-17.02,3.02) among 6 VOIs, 2.85$\mathrm{\pm}$5.21 in lesions & Dong \textit{et al.} 2020\cite{RN1826} \\ \hline 
\end{longtable}

*N.A.: not available, i.e. not explicitly indicated in the publication

${}^{+}$Numbers in parentheses indicate minimum and maximum values.

\noindent 
\section{LOW-COUNT PET RECONSTRUCTION}

Low-count PET has extensive application in pediatric PET scan and radiotherapy response evaluation with advantage of better motion control and low patient dose. However, the low count statistics would result in increased image noise, reduced contrast-to-noise ratio, and large bias in uptake measurement. The reconstruction of a standard- or full-count PET from the low-count PET cannot be achieved by simple postprocessing operations such as denoising since lowering radiation dose changes the underlying biological and metabolic process, leading to not only noise but also local uptake values changes.\cite{RN1837} Moreover, even with a same tracer injection dose, the uptake distribution and signal level can vary greatly among patients. Recently, learning-based low-count PET reconstruction methods have been proposed to take advantage of their powerful data-driven feature extraction capabilities between two image datasets. These studies are summarized in Table II. Similar as in PET AC methods, the general workflows can be divided into two groups depending on whether anatomical image is acquired (MR in PET/MR) or not (PET-only scanner). Most proposed methods were implemented on PET brain scans, with a few on lung and whole body. The ground truth for training and evaluation was full-count PET. Compared with the evaluations in PET AC which focus on relative bias, the evaluations in the reviewed studies in low-count PET reconstruction more focus on the image quality and the similarity between the predicted result and its corresponding full-count PET ground truth. The common metrics include PSNR (peak signal-to-noise ratio), SSIM (structure similarity index), CV (coefficient of variation). Several studies also compared their proposed methods with other competing learning-based methods and conventional denoising methods.

\noindent 
\subsection{Machine learning-based methods}

Learning-based methods for low dose PET reconstruction have been developed before deep learning-based methods were introduced into this topic. Kang \textit{et al.} employed a random forest to predict the full dose PET from low dose PET and MR images.\cite{RN1838} It first extracts features from patches of low-count PET and MR images based on segmentation to build tissue-specific models for initialization. Then, an iterative refinement strategy is used to further improve the predication accuracy. The method was evaluated on brain PET images with ¼ of full dose, and the quantitative evaluation on the predicted full-count PET in standard uptake value (SUV) difference from full dose PET and signal-noise-ratio showed promising results of quantification accuracy and enhanced image quality.

Wang \textit{et al.} proposed a sparse representation (SR) framework for low dose PET reconstruction.\cite{RN1839} In order to incorporate multimodal MR images in input with lose dose PET, they used a mapping strategy to ensure that the sparse coefficients, estimated from the multimodal MR images and low-count PET image, can be applied directly to the prediction of full dose PET image. An incremental refinement framework is also added to further improve the performance. A patch selection-based dictionary construction method is used to speed up the prediction process. Evaluation study on brain PET scans with ¼ of full dose demonstrated significantly improved PSNR and NMSE compared with original low-count PET. The authors also compared the performance of the proposed mapping-based SR with that of the conventional SR on the same dataset, and demonstrated that the proposed method consistently achieves higher PSNR and lower NMSE across all subjects, which indicates that the mapping strategy does help the SR enhance the prediction quality.

An \textit{et al.} also proposed to formulate full-count estimation as a sparse representation problem using a multi-level canonical correlation analysis-based data-driven scheme.\cite{RN1837} The rationale is that the intra-data relationships between the low-count PET and full-count PET data spaces are different, which result in low efficacy of direct applying learned coefficients from the L-PET dictionary to the S-PET dictionary for estimation. Canonical correlation analysis is then used to learn global mapping with the original coupled low-count PET and full-count PET dictionaries and then map both kinds of data into their common space. The multi-level scheme can further improve the learning of the common space by passing on the low-count PET dictionary atoms with non-zero coefficients of patches from first level to the next level with the corresponding full-count PET dictionary subset, rather than immediately estimating the target full-count PET at the first level. This framework is capable of using PET alone or combination of PET and multimodal MR images as input. The evaluation study on brain low-count PET patients showed that the proposed method successfully improved the visual quality and quantification accuracy of SUV, and significantly outperformed the competing learning-based methods such as SR and random forest as well as the conventional denoising methods such as BM3D\cite{RN1849} and Optimized Blockwise Nonlocal Means (OBNM)\cite{RN1850}.

Wang \textit{et al.} pointed out that the performance of SR approach depends on the completeness of the dictionary, which means that it typically requires coupled samples in the training dataset.\cite{RN1840} However, for low-count PET reconstruction with multimodal MR as additional input, it is hardly to fulfill this requirement, i.e. each sample in training has the complete modalities. They then proposed to use an efficient semi-supervised tripled dictionary learning method to effectively utilize all the available training samples including the incomplete ones to predict the full-count PET. The proposed framework enforced node-to-node and edge-to-edge matching between the patches of low-count PET/MR and full-count PET, which can reduce the requirements on their similarity. They validated the proposed method on 18 brain PET patients, only 8 among of which had complete datasets (i.e. having all low-count PET, full-count PET and MR images of three modalities). It is seen that the proposed method outperformed traditional SR and random forest in PSNR and NMSE. It also compared the performance of the proposed method with and without the incomplete datasets included in the training datasets, and the results suggested that the addition of incomplete datasets benefited the performance rather than harmed it, and should be used if available.

\bigbreak

\noindent 
\subsection{Deep learning-based methods}

The sparse-learning-based methods reviewed above usually include several steps such as patch extraction, encoding, and reconstruction. It would be time-consuming when testing new cases since it involves a large number of optimization problems, which might not be appropriate for clinical practice. Secondly, these methods tend to over-smoothen the image due to its patch-based processing, which would limit its clinical usability in fine structures detection.
	
Xiang \textit{et al.} proposed a deep auto-context CNN to predict full-count PET image based on local patches in low-count PET and MR images.\cite{RN1841} This regression method integrated multiple CNN modules by the auto-context strategy in order to iteratively improve the estimated PET image. A basic four-layer CNN was first used to build a model to estimate the full-count PET from low-count PET and MR images. The estimated full-count PET was treated as the source of the context information, and was used as inputs along with the low-count PET and MR images to a new four-layer CNN. Thus, the multiple CNNs were gradually concatenated into a deeper network, and were optimized altogether with back-propagation. Validations on brain PET/MR datasets showed the proposed method can provide competitive quality of full-count PET as the SR-based method proposed by An \textit{et al.} \cite{RN1837} with much less time in prediction. A potential limitation of this study is that the axial slices extracted from the 3D images were treated as separate 2D images independently for training the deep architecture. This would cause the loss of information in sagittal and coronal directions and discontinuous estimation results across slices.\cite{RN1850} Its patch-based workflow also tends to ignore the global spatial information in the prediction results.

Recent studies have shown that low-noise training datasets  are unnecessary for CNN to produce denoised image since CNN architecture has an intrinsic ability to solve inverse problems.\cite{RN1862} The proposed deep image prior approach iteratively learns from a pair of random noise and corrupted image, and a denoised image is then obtained as output with moderate iterations.\cite{RN1861} Hashimoto \textit{et al.} applied the deep image prior approach in CNN for dynamic brain PET imaging. The static PET images that acquired from the start to end were used as network input, and the dynamic PET images were used as label. They reported that the proposed method maintained the CNR and outperform the conventional model-based denoising methods on a dynamic brain PET study of monkey.

To fully utilize the 3D information from image volume, Wang \textit{et al.} proposed an end-to-end framework based on 3D conditional GAN to predict full-count PET from low-count PET. They used both convolutional and up-convolutional layers in the generator architecture to ensure the same size of the input and output. The generator network is a 3D Unet-like deep architecture with skip connection strategy, which is more efficient than voxel-wise estimation. Compared with 2D axial-based conditional GAN, the proposed 3D scheme was shown to produce better visual quality with sharper texture on sagittal and coronal views. Compared with the SR-based methods and CNN-based method mentioned above (Ref\cite{RN1839, RN1840, RN1841}), the proposed method featured better spatial resolution and quantitative accuracy. To incorporate multimodal MR images into input for better performance, Wang \textit{et al.} proposed an auto-context-based locality adaptive multi-modality GANs (LA-GANs) model to synthesize the full-count PET image from both the low-count PET and the accompanying multimodal MRI.\cite{RN1843} The locality adaptive means that the weight of each imaging modality depends on the image locations. 1Ã—1Ã—1 kernel was used to learn such locality-adaptive fusion mechanism to minimize the increase of the number of parameters caused by the input of multi-modality images. Compared with their previous method, it can achieve better performance with better PSNR while incurring smaller number of additional parameters.

Lei \textit{et al.} proposed a CycleGAN model to estimate diagnostic quality PET images using low count data and tested it on whole body PET scan. CycleGAN learns a transformation to synthesize full-count PET images using low-count PET that would be indistinguishable from our standard clinical protocol. The algorithm also learns an inverse transformation such that cycle low-count PET data (inverse of synthetic estimate) generated from synthetic full-count PET is close to the true low-count PET. This approach improves the GAN modelâ€™s prediction and uniqueness of the synthetic dataset. Residual blocks are also integrated to the architecture to better capture the difference of low- and full-count images and enhance convergence. Evaluation studies showed that the average bias among all 10 hold-out test patients was less than 5\% for all selected organs, and the comparison with Unet and GAN-based methods indicated a higher quantitative accuracy and PSNR in all selected organs. 

In addition to the direct mapping from low-count PET to full-count PET, some studies investigated the feasibility of embedding the deep learning network into the conventional iterative reconstruction framework. Kim \textit{et al.} modified the denoising CNN and trained with patient datasets of full-count PET and low-count PET with a preset noise level by simulation.\cite{RN1844} The predicted results from denoising CNN then serve as prior in the iterative reconstruction setting with local linear fitting function to correct the unwanted bias caused by the noise level changes in each iteration. Gong \textit{et al.} used a CNN trained with pairs of high-dose PET and iterative reconstructions of low-count PET. \cite{RN1851} Then, instead of directly feeding a noisy image into CNN and estimating its output, they used the CNN to define the feasible set of valid PET images as the initial images in iterative reconstruction. Both methods compared with conventional denoising and iterative reconstruction algorithms and demonstrated advantages in less noise when quantitative accuracy is kept same. In contrast to these methods that CNN serves as the regularizer in the reconstruction loop, Häggström \textit{et al.} exploited the deep convolutional encoder-decoder network to take the low-count PET sonogram as input and directly output the high-dose PET images thus the network implicitly learns the inverse problem. On simulated whole-body PET images, the proposed method outperformed the conventional OSEM method in more preserved fine details, less noise and less reconstruction time.

\bigbreak

\noindent 
\subsection{Discussion}

As categorized above, the reviewed studies can be mainly divided into machine learning-based and deep learning-based approaches. Although the above machine learning-based methods showed good estimation performance, most of them have limitations since they depend on the handcraft features extracted from prior knowledge. Handcrafted features have limited ability in effectively representing images. The voxel-wise estimation strategy usually involves solving a lot of optimization problems online, which is also time-consuming when applying on new test datasets. Moreover, the small-patch-based methods would average the overlapped patches as the last step, thus it tends to generate over-smoothed and blurred images, resulting in loss of texture information in PET images. In contrast, the deep learning-based methods can learn representative features directly from the training datasets. The recent advances in GPUs also allow the deep learning-based methods to be efficiently implemented with large number of parameters and trained with full-sized images. To speed up the convergence and avoid overfitting in training, techniques such as batch normalization, additional bias and ReLU have been integrated into the network. Compared to CNN which has one generator network, GAN has a generator network and a discriminator network. The purpose of the additional discriminator is to learn to distinguish between the prediction from generator and the real inputs, while the generator is optimized to predict samples such that they are indistinguishable from real data by the discriminator. The two networks are trained alternatively to respectively minimize and maximized an objective function.

Comparison among several reviewed methods has been included in a few studies. For example, Wang \textit{et al.}\cite{RN1842, RN1843} compared SR-based, dictionary learning-based and CNN-based methods \cite{RN1837, RN1839, RN1840, RN1841} with their proposed GAN-based methods. It is seen that the predicted full-count PET images by SR-based methods are more likely to generate over-smoothed images than CNN- and GAN-based methods. GAN-based methods are better than CNN-based method in preserving the detailed information, and it is argued by the authors that it is because that the CNN-based method does not consider the varying contributions across image locations. The proposed GAN-based method also outperforms other competing methods in quantification metrics of image quality and accuracy with statistical significance. The CNN-based and GAN-based methods also feature high processing speed around several seconds, as compared to minutes and even hours of machine learning-based methods. Note that these comparison studies were from the same group that also proposed the comparing methods. Independent studies in the performance evaluation of different methods are encouraged. Lu \textit{et al.}\cite{RN1848} compared the performance of CAE, Unet and GAN on lung nodule quantification application. They reported that Unet and GAN are slightly better than CAE with less difference between the predicted results and the ground truth, and the prediction of CAE is smoother. Quantitatively, Unet and GAN have significantly lower SUV bias than CAE, especially for small nodules. Unet and GAN have comparable performance while Unet has less computation cost.

The effectiveness of 3D model over 2D model is confirmed in a few publications.\cite{RN1848, RN1842} In 2D model, the whole 2D slices, either at sagittal, coronal or axial view, are used as input, while 3D model uses large 3D image volume patches as input. Results showed that the advantage of 3D model is the better performance in all the three views, while 2D model has obvious discontinuity and blur in the views different from the view of training, and substantial underestimation in quantification.

The benefit of multimodal MRI has also been indicated by the reviewed researches. Wang \textit{et al.} compared the performance of their proposed method using T1-weighted and FA and MD from DTI brain MR images jointly or separately.\cite{RN1840} They found that improvement over using low-count PET alone was achieved by including any channel of MR images as inputs. Using all multimodal MR images with low-count PET jointly can further enhance the performance. It is also seen that the general results by single modal MR image is similar to each other across all three channels, which suggests that these channels contribute similarly in most of the brain areas. Similar conclusions were also drawn in other studies with slight difference that T1-weighted may be more helpful than DTI in full-count PET prediction.\cite{RN1843, RN1840} It can be because the T1-weighted image can show both white matter and grey matter more clearly. For the DTI image, it contributes more in the white matter regions than in the grey matter regions.

Lei \textit{et al.} presented the first study of learning-based low-count PET reconstruction on whole body. Similar as the cases in PET AC studies, the application of these data-driven approaches to brain data are often less complicated due to less inter-patient anatomical variation in brain images as compared to that on whole body images. Whole body PET images also have higher intra-patient uptake variation, i.e. tracer concentration is much higher in brain and bladder than anywhere else, which may decrease the relative contrast among other organs and introduce difficulty in extracting features. Moreover, considering the challenges in PET/MR on whole body, the benefit from MR images in full-count prediction are worthy of re-evaluation.

\noindent Table II. Overview of learning-based low-count PET reconstruction methods.

\begin{longtable}{|p{0.6in}|p{0.7in}|p{0.5in}|p{0.9in}|p{0.4in}|p{1.1in}|p{0.8in}|} \hline 
Methods and strategy & PET or PET+other modalities & Site & \# of patients in training/testing datasets & Count fraction (low-count/ full-count) & Reported image quality metrics${}^{+}$: low-count/predicted full-count/full-count & Authors \\ \hline 
Random forest & PET+MR(T1-weighted) & Brain & 11, leave-one-out & 1/4 & Coefficient of variation\newline 0.38/0.33/0.31 & Kang \textit{et al.} 2015\cite{RN1838} \\ \hline 
Multi-level canonical correlation analysis & PET, PET+MR(T1-weighted, DTI) & Brain & 11, leave-one-out & 1/4 & PSNR:\newline 19.5/23.9/N.A.* & An \textit{et al.} 2016\cite{RN1837} \\ \hline 
Sparse representation & PET+MR(T1-weighted, DTI) & Brain & 8, leave-one-out & 1/4 & PSNR:\newline 19.02 (15.74, 22.41)/19.98 (16.92, 23.24)/ N.A. & Wang \textit{et al.} 2016\cite{RN1839} \\ \hline 
Dictionary learning & PET+MR(T1-weighted, DTI) & Brain & 18, leave-one-out (only 8 used for test) & 1/4\newline  & PSNR${}^{\#}$:\newline (16, 20)/(24, 27)/ N.A. & Wang \textit{et al.} 2017\cite{RN1840} \\ \hline 
CNN & PET or PET+MR(T1- and T2-weighted) & Brain & 40, five-fold cross-validation ${}^{18}$F-florbetaben & 1/100 & PSNR${}^{\#}$: 31/36(PET-only), 38(PET+MR)/ N.A. & Chen \textit{et al.} 2019\cite{RN1846} \\ \hline 
CNN & PET & Brain, lung & Brain: 2 training/1 testing\newline Lung: 5 training/1 testing & Brain: 1/5\newline Lung: 1/10 & N.A. & Gong \textit{et al.} 2019\cite{RN1860} \\ \hline 
CNN (without training datasets) & PET & Brain & 0 training/ 1 test (monkey) & 1/5 and 1/10 & CNR:\newline N.A./10.74 (1/5 counts), 7.44 (1/10 counts)/11.71 & Hashimoto \textit{et al.} 2019\cite{RN1861} \\ \hline 
Deep auto-context CNN & PET+MR (T1-weighted) & Brain & 16, leave-one-out & 1/4 & PSNR:\newline N.A./24.76/ N.A. & Xiang \textit{et al.} 2017\cite{RN1841} \\ \hline 
Iterative Reconstruction using Denoising CNN & PET & Brain & 27 training (${}^{11}$C-DASB), 1 testing & 1/6 & SSIM: N.A./0.496/ N.A. & Kim \textit{et al.} 2018\cite{RN1844} \\ \hline 
Iterative Reconstruction using CNN & PET & Brain, lung & Brain: 15 training/1 validation/1 testing\newline Lung: 5 training/1 testing & 1/10 & N.A. & Gong \textit{et al.} 2019\cite{RN1851} \\ \hline 
Deep convolutional encoder-decoder  & PET (sinogram) & Whole body (simulated) & 245 training, 52 validation, 53 testing & N/A & PSNR: 34.69 & H\"{a}ggstr\"{o}m \textit{et al.} 2019\cite{RN1852} \\ \hline 
GAN & PET & Brain & 16, leave-one-out & 1/4 & PSNR${}^{\#}$:\newline 20/23/ N.A. (normal)\newline 21/24/ N.A. (mild cognitive impairment)\newline  & Wang \textit{et al.} 2018\cite{RN1842} \\ \hline 
GAN & PET+MR(T1, DTI) & Brain & 16, leave-one-out & 1/4 & PSNR:\newline 19.88$\mathrm{\pm}$2.34/24.61$\mathrm{\pm}$1.79/ N.A. (normal), \newline 21.33$\mathrm{\pm}$2.53/25.19$\mathrm{\pm}$1.98/ N.A. (mild cognitive impairment)\newline  & Wang \textit{et al.} 2019\cite{RN1843} \\ \hline 
GAN & PET & Brain & 40, four-fold cross validation & 1/100 & PSNR${}^{\#}$:\newline 24/30/ N.A. & Ouyang \textit{et al.} 2019\cite{RN1847} \\ \hline 
Comparison of CAE, Unet and GAN & PET & Lung & 10, five-fold cross validation & 1/10 & N.A. & Lu \textit{et al.} 2019\cite{RN1848} \\ \hline 
CycleGAN & PET & Whole body & 25 leave-one-out + 10 hold-out tests & 1/8 & PSNR:\newline 39.4$\mathrm{\pm}$3.1/46.0$\mathrm{\pm}$3.8/ N.A. (leave-one-out)\newline 38.1$\mathrm{\pm}$3.4/41.5$\mathrm{\pm}$2.5 (hold-out) & Lei \textit{et al.} 2019\cite{RN1707} \\ \hline 
\end{longtable}

*N.A.: not available, i.e. not explicitly indicated in the publication.

${}^{\#}$Estimated from figures.

${}^{+}$Numbers in parentheses indicate minimum and maximum values.

\bigbreak

\noindent 
\section{SUMMARY AND OUTLOOK}

Recent years have witnessed the trend that machine learning, especially deep learning, is being increasingly used in the application of PET imaging. Various types of machine learning networks have been borrowed from computer vision field and adapted to specific clinical tasks for PET quantitative imaging. As reviewed in this paper, the most common applications are PET AC and low-count PET reconstruction. It is also an emerging field since all of these reviewed studies were published within five years. With the development in both machine learning algorithm and computing hardware, more learning-based methods are expected to facilitate the clinical workflow of PET imaging with more potential quantification application.

In addition to PET AC and low-count reconstruction, there are other topics in PET imaging where machine learning can be exploited. For example, high resolution PET has great potential in visualizing and accurately measuring the radiotracer concentration in structures with dimensions of millimeter, while it is subject to the partial volume effect due to the limited spatial discriminating ability of scanner.\cite{RN1854, RN1853} In addition to the advancement in state-of-art PET detector that achieves sub-nanosecond coincident time resolution and sub-millimeter coincident spatial resolution,\cite{RN1853, RN1771, RN1773} the partial volume effect caused by the residual spatial blurring can be further suppressed by image-based correction methods with the aid of a second image which has substantially reduced partial volume effect such as MR and CT.\cite{RN1778, RN1774, RN1775, RN1776, RN1777} Machine learning can be promising since it has excellent performance of direct end-to-end mapping shown in other applications. Encouraging results have been shown by Song \textit{et al.}\cite{RN1856} in a preliminary study using CNN. In addition to the improvement on image quality and quantification accuracy, machine learning methods are also attractive to other advanced applications such as segmentation and radiomic, especially when its success on other imaging modalities has been demonstrated.\cite{RN1790, RN1683, RN1858, RN1793, RN1803, RN1713, RN1789, RN1716, RN1804, RN1798, RN1799, RN1802, RN1801, RN1682, RN1680} Although machine learning has been developed for decades, all of these application using machine learning are still new to this field. Therefore, it is expected to see an increasing number of publications coming out in the next few years.

The reviewed studies in the paper are all feasibility studies with small to intermediate number of patients in training/testing. The clinical utility and its potential impact of these learning-based methods cannot be comprehensively evaluated until a large number of clinical datasets are involved in the study due to the data-driven property of machine learning. Moreover, the representativeness of training/testing dataset needs special attention in clinical study. The missing of diverse demographics and pathological abnormalities may reduce the robustness and generality in the performance of the proposed method. Care has to be taken when the model is trained and applied on data from different center/scanner/protocol, which may lead to unpredictable performance.

\noindent 
\bigbreak
{\bf ACKNOWLEDGEMENT}

This research was supported in part by the National Cancer Institute of the National Institutes of Health under Award Number R01CA215718 and Emory Winship Cancer Institute pilot grant.

\noindent 
\bigbreak
{\bf Disclosures}

The authors declare no conflicts of interest.

\noindent 

\bibliographystyle{plainnat}  % needs package natbib
\bibliography{PET_review}      

\end{document}